\documentstyle[pra,aps,epsf,amssymb,cite,preprint]{revtex}

\newcommand{\bra}[1]{\mathop{\langle\left.{#1}\right|}\nolimits}
\newcommand{\ket}[1]{\mathop{\left|{#1}\right.\rangle}\nolimits}
\newcommand{\braket}[2]{\mathop{\langle{#1}\left.\right|{#2}\rangle}\nolimits}
\newcommand{\ketket}[1]{\mathop{\left|\left|{#1}
\right.\right.\rangle\!\rangle}\nolimits}
\newcommand{\brabra}[1]{\mathop{\langle\!\langle\left.\left.{#1}
\right|\right|}\nolimits}

\begin{document}

\tighten  

\title{Quantum cryptography with continuous alphabet}
\author{Denis Sych\thanks{sych@comsim1.phys.msu.su}, Boris Grishanin, and Victor Zadkov }
\address{International Laser Center and Faculty of Physics,\\ M.\ V.\ Lomonosov
Moscow State University, Moscow 119899, Russia}
\maketitle

\begin{abstract}
A new quantum cryptography protocol, based on all unselected states of a
qubit as a sort of alphabet with continuous set of letters, is proposed. Its
effectiveness is calculated and shown to be essentially higher than those of
the other known protocols.
\end{abstract}

\pacs{PACS numbers: 03.67.-a, 03.67.Dd, 03.67.Mn, 03.65.Ta}

\section{Introduction}

Quantum cryptography (QC) could well be the first practical
application of the rapidly developing field of quantum information
\cite{Gisin}. Since 1970s when the idea of QC was proposed first
\cite{wiesner,Gisin} a number of different protocols implementing
it have been suggested \cite{BB84,B92,B98,GG2002}. Despite their
diversity all of them are based on a beautiful idea employing a
basic principle of quantum mechanics---no-cloning (or
impossibility to copy) quantum states \cite{nocopy}. Thanks to
this, an eavesdropper cannot intercept the quantum communication
channel without disturbing a transmitting message if it contains a
set of \emph{incompatible}, i.e., essentially quantum, not
governed by the rules of classical logic, states. Moreover, any
attempt to copy this set of states inevitably disturbs the
transmitted message.

Keeping this advantage of quantum physics for QC in mind, any QC-protocol
uses messages entirely composed of an incompatible set of quantum states or
so called \emph{quantum alphabet} that consists of incompatible ``letters''.
Various protocols of QC are distinguished in essence only by different
alphabets, which ensure secure message transmission up to an error level
that determines the protocol efficiency. Analyzing distortions in received
messages one can reveal an eavesdropping attack, but in order to
raise the information transmission rate it is necessary also to counter such attacks.
All discussed in the literature QC-protocols have relatively
low, about 15\%, quantum bit error rate (QBER) \cite{Gisin} above which they
do not ensure secure transmission. Ideally, all perturbations in the
transmitted information are caused by an eavesdropper, but in reality
imperfections of apparatus used for realization of the QC-protocol and
external sources of noise (besides the eavesdropper) also perturb the
information.

For the optimal efficiency analysis of various protocols different
efficiency criteria are used \cite{FP}, which is inconvenient for
the objective comparison of the protocols. In this paper, we will
use most appropriate criteria based on the classical Shannon
information approach estimating amount of information transmitted
through the information channels of the QC scheme \cite{Shannon}.
In QC, typical information system includes three basic components,
Alice, Bob, and Eve (the conventional names for the sender,
receiver, and eavesdropper, respectively), which communicate
through a quantum channel. Despite the communication nature
between Alice, Bob, and Eve is quantum, they in the final analysis
exchange classical information. Therefore, the classical Shannon
information is a valid measure for a quantitative analysis of the
quantum cryptography protocols, which corresponds to the joint
probability distribution of the measurement results (which are
classical) in the quantum system Alice-Eve-Bob.

An alphabet used for a message encrypting is formed commonly by selection of a
set of quantum states at the input and output of the quantum channel. The
selection rules determine different QC-protocols. For example, QC-protocol
proposed in 1992 by Bennett, hence the name B92 \cite{B92}, uses only
\emph{two} quantum states, which is the minimum limit of number of
incompatible ``letters'' composed the alphabet. The first protocol for QC
proposed in 1984 by Bennett and Brassard (BB84) \cite{BB84} gives another
example of protocol using \emph{four} quantum incompatible states.

In another limiting case, when selection of quantum states is not
performed and, therefore, the alphabet consists of \emph{all}
states of the Hilbert space, we have a new QC-protocol, which is
analyzed in this paper. It is shown that this protocol surpasses
all known QC-protocols by a number of criteria. For instance, its
critical QBER exceeds that one for the BB84 protocol and
generalization of our protocol to the case of multidimensional
Hilbert space further significantly improves the critical QBER.
This means that the new protocol can basically work at any level
of external errors or eavesdropping attacks, which is a new
feature of the QC-protocols.

\section{Compatible information as a quantum information measure in QC-protocols}

In QC, Alice ($A$), Bob ($B$), and Eve ($E$) are the different,
kinematically independent quantum systems. Thus, the quantum events related
to these systems belonging to the different Hilbert spaces are mutually
compatible. Due to this property, any pair of quantum events at the input
and output of the quantum channel can be considered classically. Quantum
specific of the channel is revealed then only in the form of intrinsic
quantum uncertainty of events at the input and output of the channel. We
will call information related to the mutually compatible events in two
quantum systems the \emph{compatible} quantum information
\cite{PIT02,RTE02}. A natural quantitative measure for the compatible
information is the standard mutual Shannon information functional of the
classical input--output (Alice-Bob) joint probability distribution $P_{AB}$:
\begin{equation}\label{qinf}
I_{AB}[P_{AB}]=S_A[P_A]+S_B[P_B]-S_{AB}[P_{AB}],
\end{equation}

\noindent where $S[P]$ is the classical Shannon entropy functional
for $P=P_A$, $P_B$, and $P_{AB}$ \cite{Shannon}.

In quantum theory, like in the classical theory of information, one has to
clarify which quantum events are used for the information exchange between
quantum systems and to define a set of elementary events of which any
message is composed. Elementary events for a given quantum system are
represented by the wave functions or state vectors of the system.
Mathematically, a choice of basis events can be given by defining a set of
positive operators $\hat E_\nu$ representing a non-orthogonal expansion of
unit operator \cite{POVM,Preskill}, which is an analog of average classical
indicator functions of a complete group of classical random events, which are
normalized to the indicator of the reliable net event (with the probability
equal to unity) and can be written in quantum case as the unit operator:
\begin{equation}\label{qp}
\hat 1 = \sum\hat E_\nu.
\end{equation}

\noindent For simplicity, we will consider in the following
two-dimensional spaces, if not defined otherwise.

Two limiting cases of the compatible information---limiting \emph{selected}
and \emph{non-selected} information---are defined by two limiting cases of
the unit operator expansion---two-component orthogonal \cite{vonNeuman}
\begin{equation}\label{ort}
\hat 1=\ket\mu\bra\mu+
\ket{\tilde\mu}\bra{\tilde\mu},
\end{equation}

\noindent and continual non-orthogonal \cite{RTE02}
\begin{equation}\label{nort}
\hat 1=\int\limits_\nu\!\! \ket\nu\bra\nu{\rm d} V_\nu,
\end{equation}
\noindent where $\ket\mu$ and $\ket{\tilde\mu}$ are the arbitrary
pair of orthogonal wave functions and ${\rm d} V_\nu= \sin\theta
d\theta d\varphi/(2\pi)$ with the  standard angular parameters on
the Bloch sphere.

The selected information determines an information link between two quantum
systems $A$ and $B$ with the joint density matrix $\hat\rho_{AB}$ through the
selected set of orthogonal quantum events. The orthogonal basis determined
by the unitary two-parametric transformation $U_A(\alpha)$ and $U_B(\beta)$
in quantum systems $A$ and $B$, respectively, can be chosen differently and
the selected information also depends on the choice made:
\begin{equation}\label{sinf}
\displaystyle I_{AB}(\alpha,\beta) = \sum\limits_{k,l}
P_{AB}^{\alpha\beta}(k,l)
\log_2\frac{P_{AB}^{\alpha\beta}(k,l)}{P_A^\alpha(k)P_B^\beta(l)},
\end{equation}

\noindent where parameters $\alpha=(\theta_1,\varphi_1)$ and
$\beta=(\theta_2,\varphi_2)$ are given by the standard Bloch sphere angles.

Joint distribution $P_{AB}^{\alpha\beta}(k,l)={\rm Tr}_{AB}\bigl(\hat
E_A^\alpha(k) \otimes \hat E_B^\beta(l)\bigr) \hat\rho_{AB}$ is
defined on the basis states of the input (Alice) and output (Bob)
of the channel $\ket{k}_A^\alpha$, $\ket{l}_B^\beta$ numbered by
two indices $k$ and $l$. These states are the orthogonal basis
states of respected Hilbert spaces $H_A$ and $H_B$.

For the non-selected information, the information exchange undergoes through
all equally participating in the exchange states. In this case, basis states
of the information channel are \emph{all} wavefunctions of the Hilbert
spaces of all participating in the exchange quantum systems. The respected
non-selected information is then given as
\begin{equation}\label{uinf}
I_{AB}=\int\limits_\alpha\!\!\!\int\limits_\beta\!\!
P_{AB}({\rm d}\alpha,{\rm d}\beta)
\log_2\frac{P_{AB}({\rm d}\alpha,{\rm d}\beta)}{P_A({\rm d}\alpha)P_B({\rm d}\beta)},
\end{equation}

\noindent where $P_{AB}({\rm d}\alpha,{\rm d}\beta)={\rm Tr}_{AB}\bigl(\hat
E_A({\rm d}\alpha) \otimes \hat E_B({\rm d}\beta)\bigr)\hat\rho_{AB}$,
$\hat E_A({\rm d}\nu)=\ket\nu_A\bra\nu_A {\rm d} V_\nu$.

Note that the non-selected information is equal to the selected one, which
is averaged over all orientations of the orthogonal bases:
\begin{equation}\label{suinf}
I_{AB}=\int\limits_\alpha\!\!\!\int\limits_\beta\!\!
I_{AB}({\rm d}\alpha, {\rm d}\beta)\frac{{\rm d} V_\alpha {\rm d}
V_\beta}{V^2},\quad V=\int\!\! {\rm d} V_\nu=2.
\end{equation}

\section{QC-protocol employing all states of the Hilbert space of the
cryptographic system}

Key idea of QC is that secure quantum information channel is used
first to transfer a secret key from Alice to Bob, which is then
used to encrypt messages transmitted via an insecure classical
channel. The quantum channel over which the message out of which
the secret key will be extracted is transferred can be
eavesdropped by Eve who can perform any physically allowed
transformations gaining information about the transferring
message. The purpose of Alice and Bob is to establish a secure
connection, which prevents copying of useful transmitted
information by Eve. Ii was proved that such secure connection is
possible if the amount of information Bob received from Alice
exceeds information Eve received either from Alice or Bob
\cite{coding}. This condition can be written as
\begin{equation}\label{opt}
I_{AB}>{\rm max}\bigl(I_{AE},I_{BE}).
\end{equation}

\noindent If the condition (\ref{opt}) is fulfilled, it is possible with the
help of special methods of encrypting/decrypting the information to reduce up
to zero the amount of useful information Eve can gain eavesdropping the quantum
channel.

Eve, in her turn, also tries to use optimal strategies of eavesdropping at
the given level of interference, i.e., Eve tries to gain maximum information
about the transmitting message minimizing the error level she causes.

All known QC-protocols using finite-dimensional spaces of states are built
on the alphabets with the finite discrete set of incompatible quantum
``letters'', which can be realized as the pure states of a quantum system. In
this paper, we suggest a qualitatively new QC-protocol, which is based on
the alphabet including all states of the Hilbert space. In other words, this
alphabet consists of an infinite number of quantum ``letters'', which are
formed by arbitrary superpositions of orthogonal basis states of the Hilbert
space $H_A$.

Let us first consider the case of two-dimensional space
(multidimensional case is considered in section
\ref{sec:multidimensional}).

Elementary step of the QC-protocol that is transmission of a single
``letter'' or state from Alice to Bob can be outlined as follows:
\begin{enumerate}
\item Alice transmits to Bob via a quantum channel a randomly
chosen state $\ket\beta$. Let us assume that the Alice's state $\ket\alpha$
is totally entangled with the transmitting state $\ket\beta$ and is, for
example, the antisymmetric Bell state
$\ket{\ket-}=\bigl(\ket\alpha\ket{\tilde\beta}- \ket{\tilde\alpha}
\ket\beta\bigr)/\sqrt2$, which means that Alice perfectly knows the
transmitting state.

\item Eve eavesdrops the channel performing an unitary
transformation $U_{BE}$ with her initial state $\ket0_E$ and with
transmitted by Alice to Bob state $\ket{\beta}_B$ and readily
measures her final state.

\item Bob reads the perturbed state using for the measurement an
\emph{arbitrary} projector because he has no \emph{a priori} information
about the received message.
\end{enumerate}

When transmission of the message is completed, Alice and Bob
disclose part of the measurement results transmitting them over an
insecure classical channel in order to determine the mutual
probability distribution, which is then used for calculation of an
average amount of transmitted information per an elementary step
of the QC-protocol. After that, disclosed results are discarded
and not used for further generation of a secret key. If the
security condition (\ref{opt}) is fulfilled, Alice and Bob decide
that the secret key transfer is completed, otherwise the
transmitted key is not used. Note that the described protocol does
not require bases reconciliation of Alice and Bob, i.e., selection
of only that part of the message for which Alice and Bob used the
same measurement projectors, via an additional information
exchange over the classical channel.

\subsection{Information analysis of the protocol}

For information analysis of the suggested protocol let us first calculate
the amount of information Bob received from Alice and Eve received from
Alice and Bob at the condition of optimal eavesdropping.

Initial state of the quantum system Alice-Eva-Bob
$\hat\rho_{ABE}^{(1)}= \hat \rho_{AB}^{(1)} \otimes
\ket0_E\bra0_E$, which is described by the tensor product of the
entangled antisymmetric pair Alice-Bob $\hat\rho_{AB}^{(1)}=
\ketket{-}_{AB} \brabra{-}_{AB}$ and an initial Eve's state
$\ket0_E$, is transferred after eavesdropping the channel by Eve
into the final state that is an entangled state of Alice, Bob, and
Eve, $\hat\rho_{ABE}^{(2)}$:
$\hat\rho_{ABE}^{(1)}\stackrel{U_{BE}}{\longrightarrow}
\hat\rho_{ABE}^{(2)}$.

We can then assume (without reducing the generality of our
consideration) that the unitary transformation $U_{BE}$ performed
by Eve has the form:
\begin{equation}\label{ueve}
\left\{\begin{array}{l} \ket0_B\ket0_E
\stackrel{U_{BE}}{\longrightarrow}
\ket0_B\ket{\Phi_{00}}_E+\ket1_B\ket{\Phi_{01}}_E\\
\ket1_B\ket0_E \stackrel{U_{BE}}{\longrightarrow}
\ket0_B\ket{\Phi_{10}}_E+\ket1_B\ket{\Phi_{11}}_E.
\end{array}\right.
\end{equation}

\noindent The unitarity imposes the following restrictions, which
are due to the orthogonality and normalization conditions:
\begin{equation}\label{cond}
\braket{\Phi_{00}}{\Phi_{10}}+\braket{\Phi_{01}}{\Phi_{11}}=0,\quad
|\Phi_{00}|^2+|\Phi_{01}|^2=|\Phi_{10}|^2+|\Phi_{11}|^2=1.
\end{equation}

It was suggested in reference \cite{FP} based on the numerical
estimations and then proved in reference \cite{optimalcopy} that
in the QC protocols BB84 and B92 the Eve's state at the optimal
eavesdropping lies in two-dimensional Hilbert space. This is also
true (and can be proved) for our QC-protocol. Therefore, the
states $\ket{\Phi_{ij}}$ can be written with the help of condition
(\ref{cond}) as a superposition of two basis states:
\begin{equation}\label{state}
\overrightarrow{\ket\Phi}=
\left(\begin{array}{c}\ket{\Phi_{00}}\\ \ket{\Phi_{01}}\\ \ket{\Phi_{10}}\\
\ket{\Phi_{11}}\end{array}\right)=
\left(\begin{array}{cc}\gamma_{00}&\gamma_{01}\\ \gamma_{10}&\gamma_{11}\\
\gamma_{11}&\gamma_{10}\\
\gamma_{01}&\gamma_{00}\end{array}\right){\ket0_E\choose\ket1_E},
\end{equation}

\noindent where the transformation parameters
\[\gamma_{mn} =(-1)^{mn} \cos{(\theta-m\frac{\pi}{2})}\cos{(\varphi -n\frac{\pi}{2})}\]

\noindent are determined via two angles $\theta$, $\varphi$ on the
Bloch sphere.

Resulted bipartite density matrices Alice-Bob, Alice-Eve, and Bob-Eve
received by averaging of the three-partite density matrix over the third
system enable us to calculate the respective mutual information:
\begin{equation}\label{pinf}
\begin{array}{l}
\hat\rho_{AB}^{(2)}={\rm Tr}_E\hat\rho_{ABE}^{(2)}\rightarrow I_{AB},\\
\hat\rho_{AE}^{(2)}={\rm Tr}_B\hat\rho_{ABE}^{(2)}\rightarrow I_{AE},\\
\hat\rho_{BE}^{(2)}={\rm Tr}_A\hat\rho_{ABE}^{(2)}\rightarrow I_{BE}.
\end{array}
\end{equation}

In our QC-protocol Alice sends Bob any pure state with equal probability and
neither Bob nor Eve have an \emph{a priori} chosen basis for the
measurement, thus both Eve and Bob use each an arbitrary chosen basis. After
averaging over large number of measurements we receive due to the equation
(\ref{suinf}) that the non-selected information is exactly the information
measure for our cryptographic system.

\subsection{Calculations results}

\begin{figure}[p]
\begin{center}
\epsfxsize=6cm\epsfclipon\leavevmode\epsffile{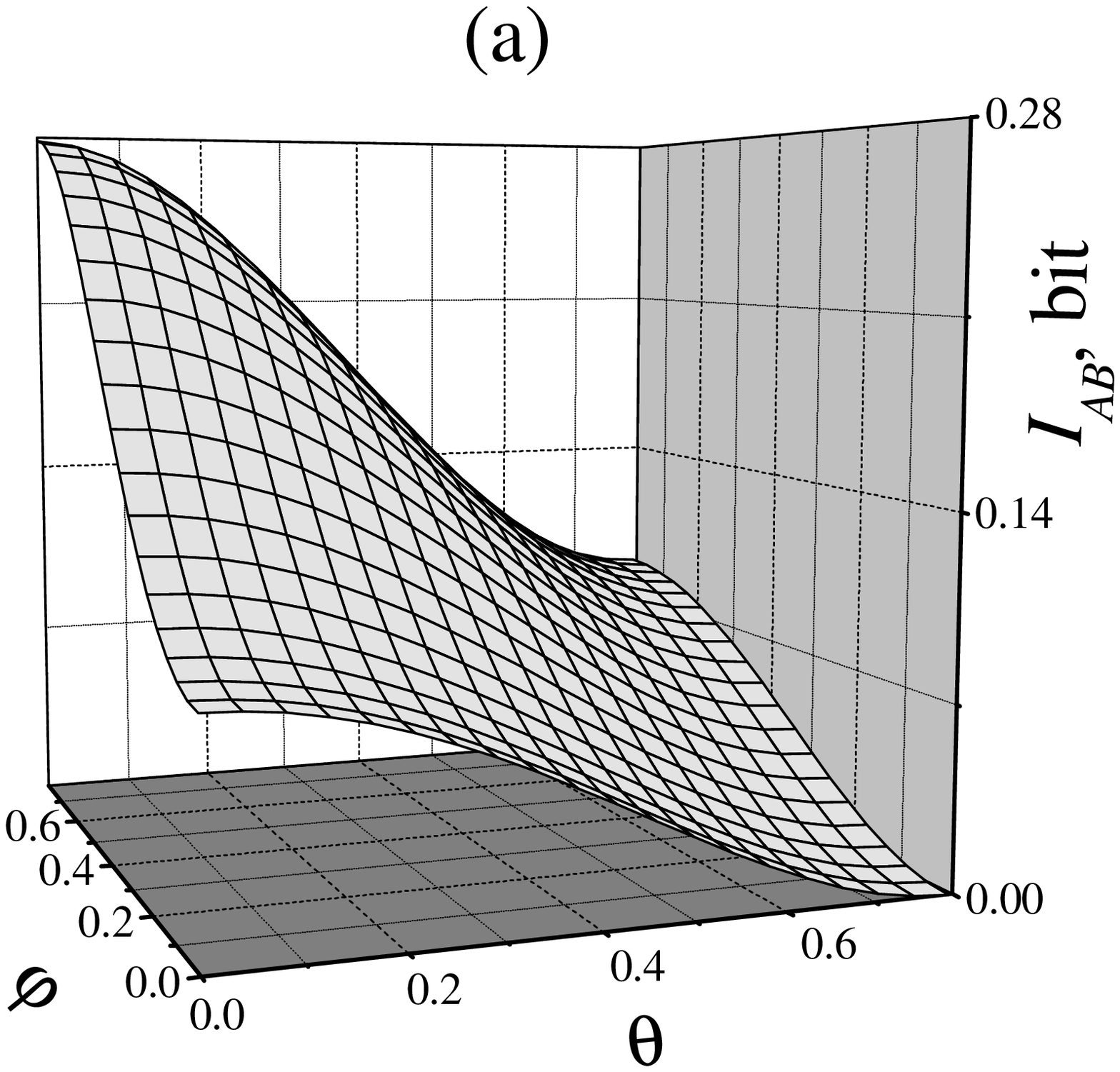}\hspace*{0.5cm}
\epsfxsize=6cm\epsfclipon\leavevmode\epsffile{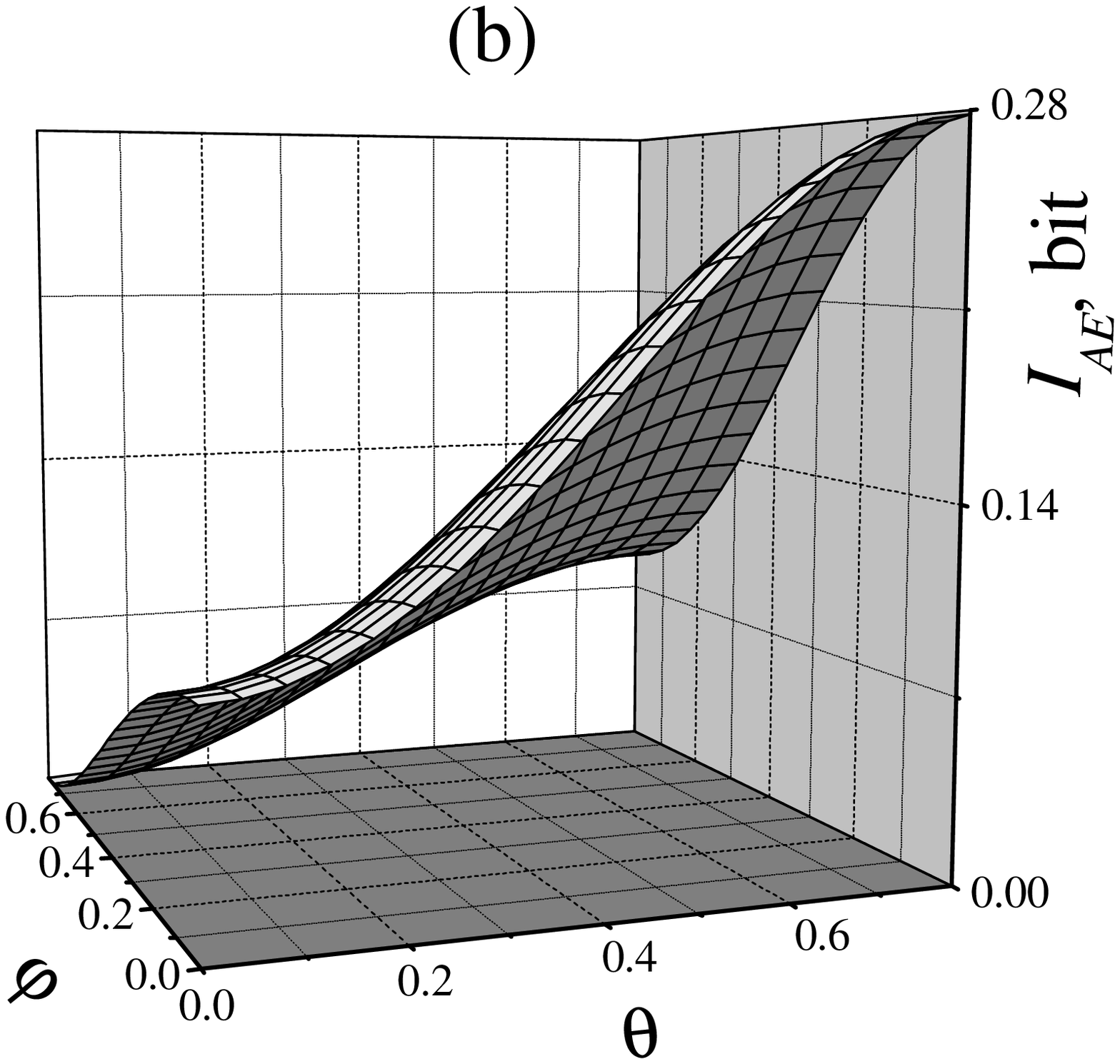}
\end{center}
\begin{center}
\epsfxsize=6.cm\epsfclipon\leavevmode\epsffile{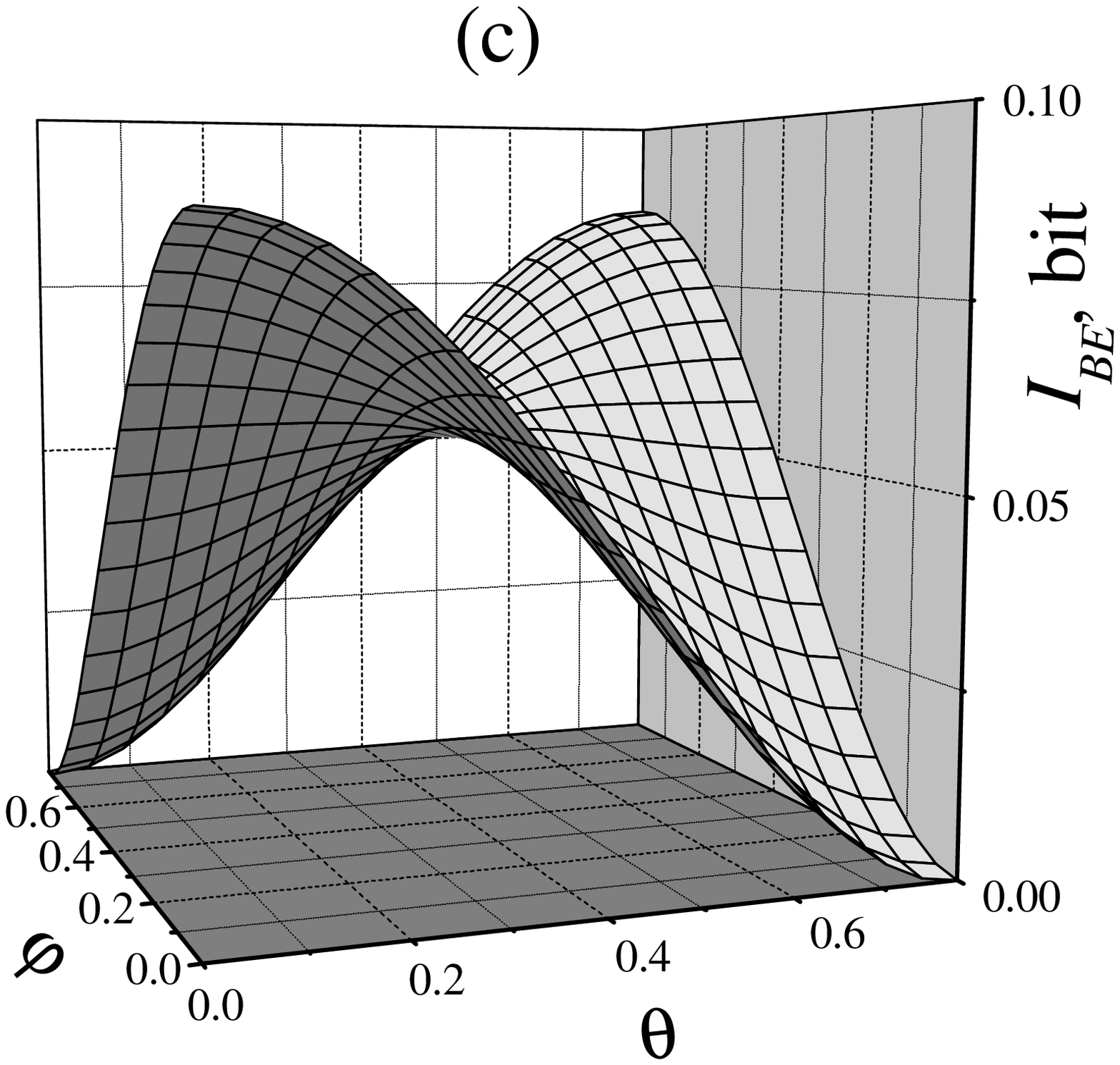}\hspace*{0.5cm}
\epsfxsize=5.5cm\epsfclipon\leavevmode\epsffile{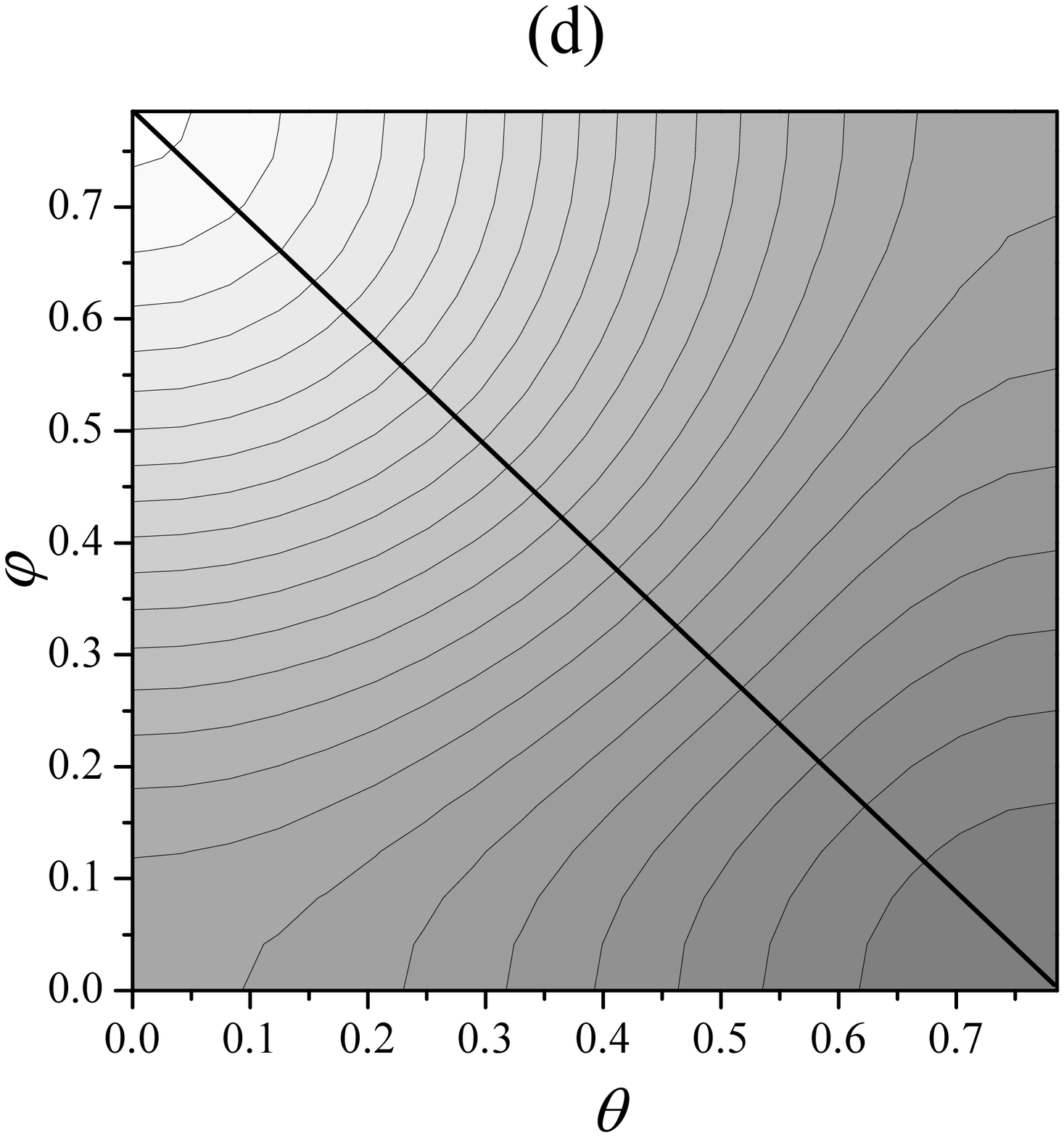}
\end{center}
\caption{Alice-Bob (a), Alice-Eve (b), and Bob-Eve (c) mutual
Shannon information versus Eve's eavesdropping parameters
$\theta$, $\varphi$. Figure (d) shows results of figure (a) for
the Alice-Bob mutual Shannon information ($I_{AB}$) as a contour
plot; solid line indicates the optimal eavesdropping.}
\label{fig1}
\end{figure}

Results for the mutual Alice-Bob, Alice-Eve, and Bob-Eve non-selected
information ($I_{AB}$, $I_{AE}$, and $I_{BE}$, respectively) calculated with
the help of equations (\ref{uinf}), (\ref{ueve}), (\ref{state}), and
(\ref{pinf}) are shown in figure \ref{fig1} versus parameters $\theta$ and
$\varphi$ controlled by Eve. One can clearly see from the figure that for
all values of $\theta$, $\varphi$ we have $I_{AE}\geq I_{BE}$, thus we will
focus in the following only on $I_{AE}$.

The optimal eavesdropping condition requires that we look for the
maximal $I_{AE}=I_{AE}(\theta,\varphi)$ at the given value of
$I_{AB}=I_{AB}(\theta,\varphi)$. Detailed analysis of data in figure
\ref{fig1} shows that the optimal eavesdropping at any level $I_{AB}$
can be realized at $\theta =\pi/4-\varphi$, which corresponds to the
solid line in figure \ref{fig1}d.

\begin{figure}
\begin{center} \epsfxsize=6.cm\epsfclipon\leavevmode\epsffile{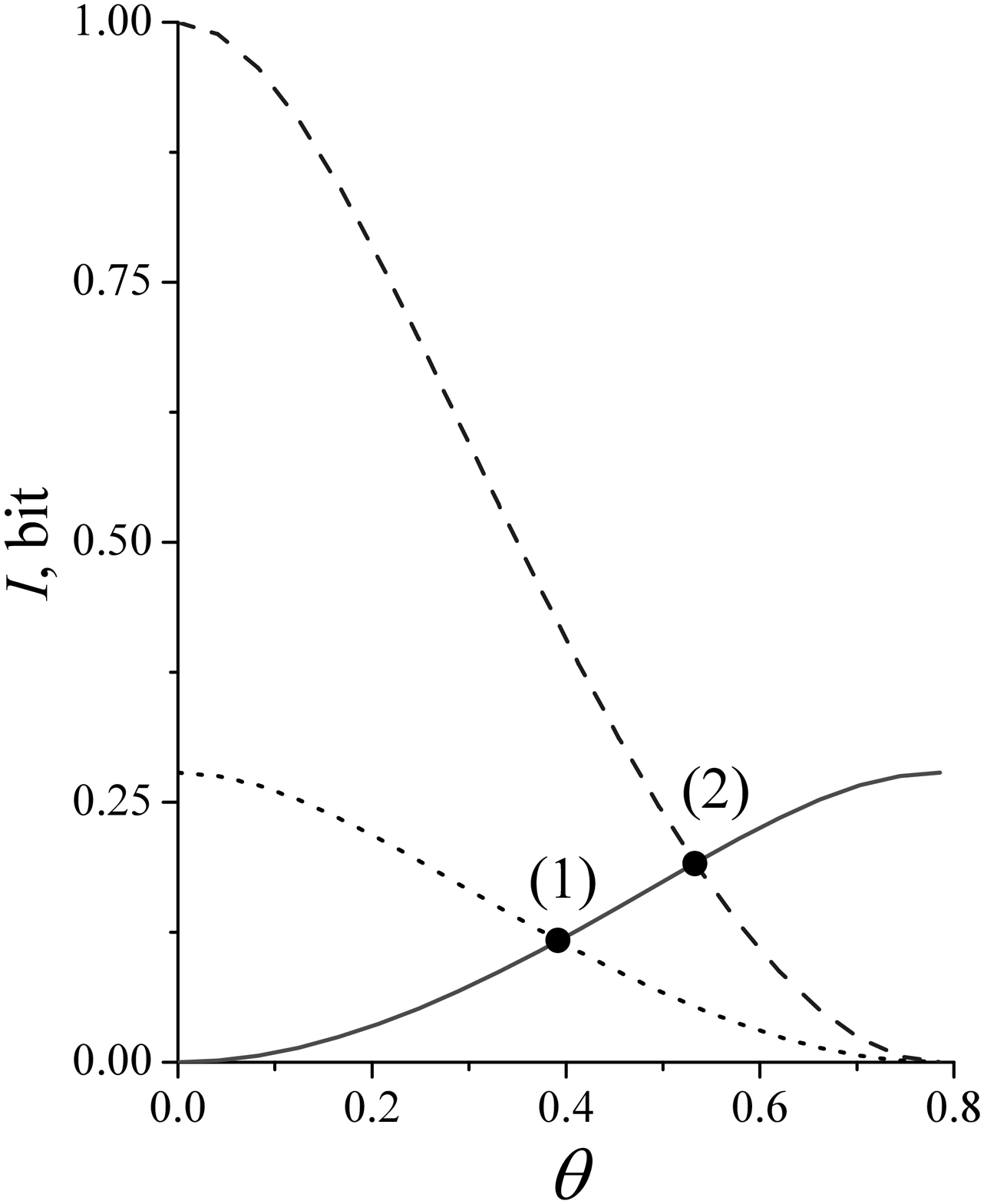}
\end{center} \caption{Alice-Bob (dotted and dashed lines for reconciliated
and non-reconciliated basis states of Alice and Bob, respectively)
and Alice-Eve (solid line) mutual Shannon information at the
optimal eavesdropping condition.}\label{fig2}
\end{figure}

For most purposes it is enough to consider only two-dimensional plots of
$I_{AB}(\theta)$ and $I_{AE}(\theta)$ along the solid line $\theta
=\pi/4-\varphi$, which are shown in figure \ref{fig2}. From analysis of this
figure one can see that at $\theta=0$ the level of Eve eavesdropping attacks
and the respected losses of information are equal to zero. At $\theta=\pi/4$
the intervention of Eve is maximal and she acts, in fact, as Bob gaining
maximal possible information.

The condition for the protocol to be a secure one, $I_{AB}>I_{AE}$, is
fulfilled up to a certain critical value $\theta_0=\pi/8$, which is the
intersection point (1) of the curves for $I_{AB}$ and $I_{AE}$ in the figure
\ref{fig2}. At this point, the QBER
$q=1-{\rm Tr}\hat\rho^{(1)}_B\hat\rho^{(2)}_B$, has the critical value of $q_0=
\sin(\pi/8) \simeq 0.15$, which is equivalent to the QBER value provided by
the BB84 protocol \cite{Gisin,BB84}. The respected value of the compatible
information is equal to $I_0=I_{AB}(\theta= \theta_0)\simeq0.11$ bit.

As it is demonstrated in reference \cite{FP}, the QBER is not always an
adequate characteristic of the degree of Eve eavesdropping attacks, for
instance in the B92 protocol. Therefore, we suggest to use another
characteristic, the \emph{compatible information error rate} (CIER), which
naturally reflects in term of the compatible information the degree of Eve
interference to the transmitting information:
\begin{equation}\label{newq}
\displaystyle Q=1-\frac{I}{I_{\rm max}}\in[0,1],
\end{equation}

\noindent where $I$ is the compatible information $I_{AB}$ with the presence
of eavesdropping and $I_{\rm max}$ its maximal possible value without Eve
attacks. By contrast with QBER ($q$), CIER ($Q$) is the most adequate
parameter for the information properties of the QC-protocols.

Without Eve eavesdropping attacks, both parameters $q$ and $Q$ are equal to
zero, which means that there are no transmission errors. At the maximal
level of Eve interference with the transmitting information, we have $Q=1$
and $q=0.5$, which correspond to the maximal possible level of errors caused
by Eve. At the critical point $\theta_0$, where the amount of information
gained by Eve is equal to the amount of information received by Bob,
$Q_0\simeq0.4$ and $q_0\simeq0.15$. For our QC-protocol, the larger critical
value $Q_0$ the better stability of the protocol to the errors caused by Eve
and, therefore, better information properties. Thus, the protocol ensures the
security of transmitted data even at essential errors rate during the data
transmission. At the error level exceeding critical, i.e. at $Q>Q_0$, the
protocol does not ensure security of transmitted data and Alice and Bob
decide that the transmission is not completed.

In suggested QC-protocol the requirement of bases reconciliation for Alice
and Bob is lifted because even with no selection of quantum states made at
the Bob's end he receives about 0.28 bit for every elementary step of the
QC-protocol, if there is no external noise during data transmission.
However, one can significantly improve stability of the protocol for a noisy
quantum channel reconciling the basis states.

After transmission of all data through a noisy quantum channel
Alice and Bob can select only those transmitted data for
transmitting which they used approximately similar orthogonal
bases. In our case, the set of basis states is the continuum, thus
it is necessary to split it into several areas and count the bases
similar, if they are in the same area on the Bloch sphere. As a
result of such selection, Alice and Bob will receive without any
intrusion of Eve maximal information value of about 1 bit during
one elementary step of the QC-protocol. This can be clearly
understood because for an initial state of the Alice-Bob system in
the form of antisymmetric Bell state the mutual selected
information is equal to unity when use similar oriented bases of
Alice and Bob. One can suppose that Eve does not affect the data
selection with the reconciling bases and does not use additional
transformations after the bases have been reconciled. Then, she
gains no additional information.

Information that Bob receives from Alice after the bases
reconciliation is shown in figure \ref{fig2} (dashed line). The
critical error rate $Q$ is then significantly increased and is of
the order of value $Q_0\simeq0.81$ and for the QBER we have
$q_0\simeq0.42$, which is much better than for all known
QC-protocols.

Note that the bases reconciliation procedure significantly increases the
number of messages transmitted over an insecure classical channel, which in
the limit of infinitively small areas on which we split the continual
quantum alphabet grows up to infinity. Respectively, the number of selected
messages transmitted through a quantum channel is decreased. It is not
necessary, however, to infinitely increase the accuracy. As a rule, errors
during the data transmission have typical for a specific experimental QC
setup finite level. Therefore, for the bases reconciliation it is sufficient
to increase the accuracy according to the external conditions up to the
level that ensures the error rate less than the critical one at which the
QC-protocol guarantees the secure transmission of data.

\section{Multidimensional case}
\label{sec:multidimensional}

We can fundamentally improve the properties of suggested QC-protocol using
multidimensional Bob's and Alice's spaces ($D>2$). In multidimensional case,
the maximally possible amount of mutual selected information is equal to
$I_{\rm max}^D=\log_2 D$ and grows infinitely at $D\rightarrow\infty$.
Maximally possible amount of non-selected information is equal to the volume
of accessible information \cite{CF}:
\[I_{\rm accesible}^D=\log_2 D-\frac{1}{\ln 2}\sum\limits_{k=2}^D \frac{1}{k},\]

\noindent which in the limit $D\rightarrow\infty$ is restricted by
the value of $I^\infty\simeq0.61$ bit.

After reconciliation the Alice's and Bob's bases the amount of information
in the system Alice-Bob is given by maximally possible selected information,
whereas in the Alice-Eve system---by the maximally possible non-selected
information. Then, critical level of errors in the limit of
$D\rightarrow\infty$ is equal to unit:
\begin{equation}\label{res1}
Q_0^\infty=\lim\limits_{D\rightarrow\infty}
Q_0^D=1-\lim\limits_{D\rightarrow\infty} \frac{I_{\rm
accessible}^D}{I_{\rm max}^D}\simeq1-\lim\limits_{D\rightarrow\infty}
\frac{0.61}{\log_2 D}=1.
\end{equation}

\noindent This means that increasing the dimensionality of the Alice-Bob
system one can reach the critical error rate (QBER or CIER), which exceeds
any given value. The multidimensional spaces of Alice and Bob can be
realized using block coding when information is transmitted simultaneously
with the help of several qubits.

Fundamental advantage of the suggested QC-protocol that employs all states
of the Hilbert space is that it can work, in principle, at any
imperfections or noise in the quantum channel (either internal or external)
and has no any critical CIER value after which the protocol becomes
insecure. For any given CIER value one can calculate (from technical point
of view, calculations in the case of multidimensional space can be readily
done) the required dimensionality of the Alice-Bob space in order to meet
this value of CIER. Essentially more difficult is the question about Eve's
transformation structure to perform optimal eavesdropping in the
multidimensional case, but the outlined above result is qualitatively
correct, despite any specific structure of the Eve's transformation.

\section{Experimental setup for QC with continuous alphabet}

\begin{figure}
\begin{center} \epsfxsize=8.cm \epsfclipon \leavevmode
\epsffile{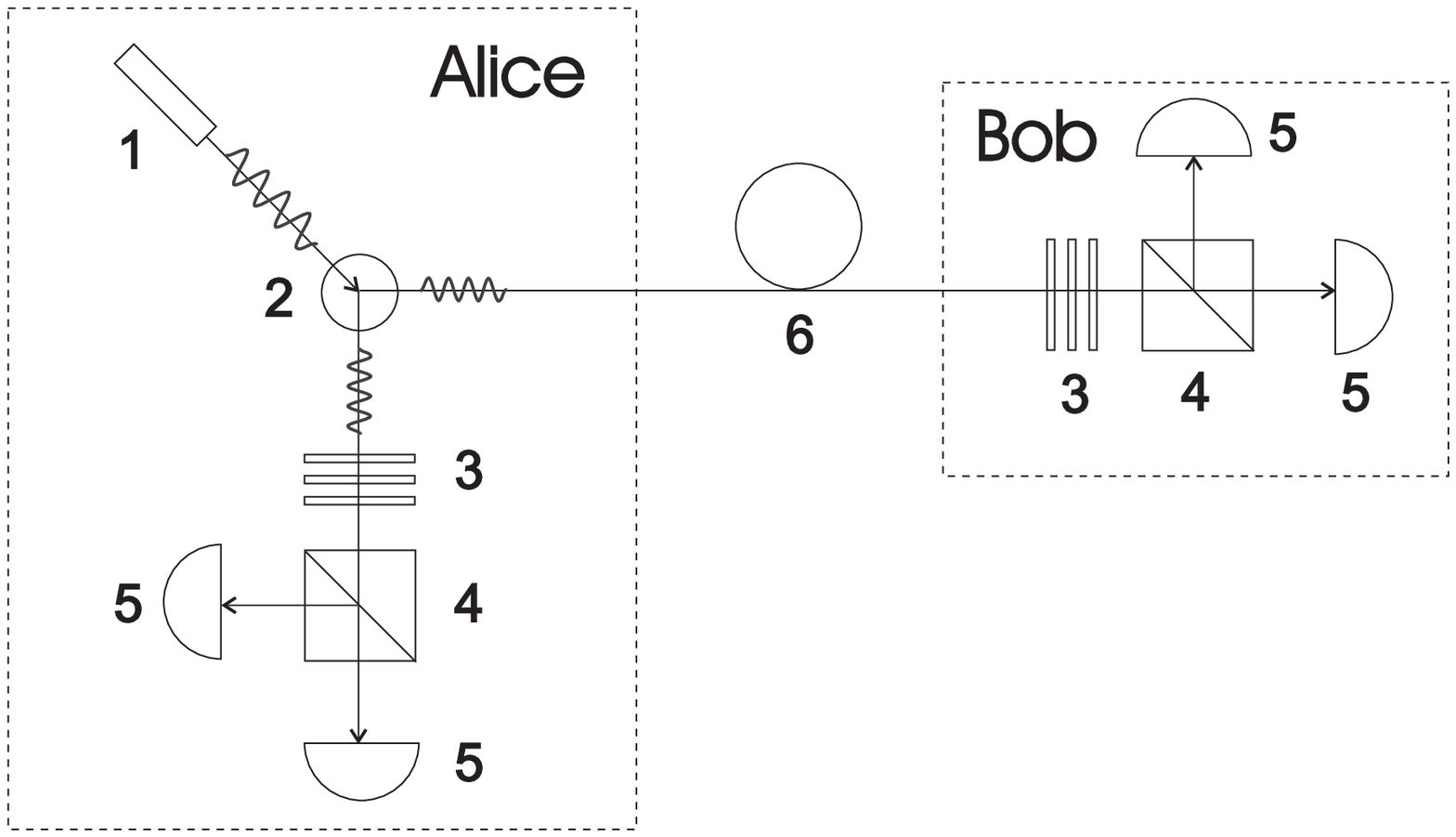}
\end{center}
\caption{Experimental setup for the QC with continuous alphabet.
Laser (1) pumps an optical parametric oscillator (2), which
generates a pair of entangled photons one of which is transmitted
then to Alice and another one, via a quantum channel (6), to Bob.
Measurement part of the cryptographic scheme consists of a
polarization plate (3) that rotates polarization of the incident
photon, prism (4), and photon counting detectors (5).}
\label{fig3}
\end{figure}

In this section, we suggest an experimental setup for the QC with
continuous alphabet ``letters'' of which are coded with
polarization of the photons (figure \ref{fig3}). Then, a random
letter corresponds to an arbitrary photon polarization.

A source of EPR-pairs of photons, for instance an optical
parametric oscillator (2) generating pairs of photons in
antisymmetric entangled state, is located at the Alice's end of
the cryptographic setup. Alice receives one of the generated
photons, another one is transmitted via a secure quantum channel
to Bob.

An arbitrary projector can be realized by rotation to an arbitrary
angle the polarization plate (3) with the following measurement in
the fixed basis. Photons transmitted from Alice to Bob are in the
antisymmetric state, thus Alice knows exactly, after the photon
measurement on her end, that photon she sent to Bob is an
orthogonal state to that one measured by Alice. As a result of the
outlined procedure, Alice sends to Bob a randomly chosen quantum
``letter''.

Likewise, Bob for the measurement in an arbitrary basis first
rotates polarization of the incident photon by polarization plate
(3) to the angle value of which he receives from Alice over an
insecure classical channel (not shown in the figure) and then
performs measurement in the fixed basis.

A case of multidimensional spaces of the quantum channel input and
output can be realized when information is transmitted with the
help of several entangled qubits (photons). This, however, is an
experimental difficulty to generate, operate, and measure
arbitrary states in multidimensional spaces, i.e., difficulty to
generate and operate multiple entangled photons.

\section{Conclusions}

In conclusion, a new QC-protocol based on the quantum alphabet with
infinitive number of ``letters'' (i.e., employing all the quantum states of
the Alice-Bob quantum system) is proposed. It has a number of advantages in
comparison with other known QC-protocols. Even in two-dimensional case the
protocol shows an essential increase of the QBER. In multidimensional case,
the protocol has a fundamental, qualitatively new feature, which allows
secure data transmission through practically any noisy quantum channel. For
estimation of the Eve's intervention into the data transmission through a
quantum channel we use classical mutual Shannon information-based criterion,
which adequately reflects the information aspect of the eavesdropping
and can be effectively used for both constructing and analyzing the
QC-protocols.

\acknowledgements

This work was partially supported by RFBR grants Nos.
01--02--16311, 02--03--32200, and INTAS Grant INFO 00--479.

\newpage

\newpage

\newpage

\end{document}